\documentclass[reprint,prx,amsmath,amssymb,aps,superscriptaddress,longbibliography]{revtex4-2}

\usepackage{color,soul}
\usepackage{graphicx}
\usepackage{dcolumn}
\usepackage{bm}
\usepackage{hyperref}
\usepackage{mathtools}
\usepackage[dvipsnames]{xcolor}
\hypersetup{colorlinks=true, linktoc=all, linkcolor=blue, linktocpage, citecolor=blue}

\newcommand{\bs}{\boldsymbol}
\newcommand{\diff}{\mathrm{d}}
\newcommand{\qs}{q_\mathrm{s}}

\newcommand{\fs}{f_\mathrm{s}}

\newcommand{\xiC}{\xi_c}

\newcommand{\kB}{k_\mathrm{B}}

\newcommand{\Eqref}[1]{\mbox{Eq.\hspace{0.25em}\eqref{#1}}}
\newcommand{\Eqsref}[1]{\mbox{Eqs.\hspace{0.25em}\eqref{#1}}}

\newcommand{\figref}[1]{\mbox{Fig.\hspace{0.25em}\ref{#1}}}
\newcommand{\Figref}[1]{\mbox{Fig.\hspace{0.25em}\ref{#1}}}

\newcommand{\Appref}[1]{\mbox{Appendix\hspace{0.25em}\ref{#1}}}

\newcommand{\mysection}[1]{}
\newcommand{\mysubsection}[1]{}

\begin{document}

\title{Phase separation with non-local interactions}

\author{Filipe C. Thewes}
\affiliation{%
 Max Planck Institute for Dynamics and Self-Organization, Am Faßberg 17, 37077 Göttingen, Germany
}%

\author{Yicheng Qiang}
\affiliation{%
 Max Planck Institute for Dynamics and Self-Organization, Am Faßberg 17, 37077 Göttingen, Germany
}%

\author{Oliver W. Paulin}
\affiliation{%
 Max Planck Institute for Dynamics and Self-Organization, Am Faßberg 17, 37077 Göttingen, Germany
}%

\author{David Zwicker}%
\affiliation{%
 Max Planck Institute for Dynamics and Self-Organization, Am Faßberg 17, 37077 Göttingen, Germany
}%

\begin{abstract}
Phase separation in complex systems is a ubiquitous phenomenon. While simple theories predict coarsening until only macroscopically large phases remain, concrete models often exhibit patterns with finite length scales. To unify such models, we here propose a general field-theoretic model that combines phase separation with non-local interactions. Our analysis reveals that long-range interactions generally suppress coarsening, 
whereas systems with non-local short-range interactions additionally exhibit a continuous phase transition to patterned phases. 
Only the latter system allows for the coexistence of homogeneous and patterned phases, which we explain by mapping to the conserved Swift--Hohenberg model.
Taken together, our generic model reveals an underlying framework that describes similar phenomena observed in 
many complex phase-separating systems. %
\end{abstract}

\maketitle

Phase separation, describing the demixing of a homogeneous mixture into dense and dilute regions, is ubiquitous in nature and materials science~\cite{Feric2016,Lafontaine2020,Berry2018,Weber2019,Alberti2017,Choi2013,Utada2005,Mao2020,Zwicker2025}.
Standard demixing is qualitatively well-described by classical theories, in particular Flory's regular solution theory~\cite{Flory1942,Huggins1941}.
One key property of such theories is that they penalize interfaces, thus favoring one large droplet over multiple smaller ones.
This leads to a coarsening process known as Ostwald ripening~\cite{Voorhees1985}, which drives systems toward macroscopically large homogeneous phases.
Yet suppressed coarsening, featuring patterns with finite length scales, has been observed for many phase separating systems with complex interactions, including block co-polymers~\cite{Ohta1986,Liu1989,Uneyama2004}, non-local elasticity~\cite{Kawasaki1988,Onuki2001-kz,Tanaka2022-la,Fernandez-Rico2022-wu,Qiang2024-li,paulin2025}, chemical reactions~\cite{Bauermann2024-qz, Christensen1996,Zwicker2022, Zwicker2015}, material exchange between compartments~\cite{Rossetto2025}, and membrane interactions~\cite{Yu2025-eo,Winter2024}. %
However, a general theory of suppressed coarsening is lacking.

Inspired by the phenomenological similarity of different systems with suppressed coarsening, we here propose that such systems can be qualitatively described by phase separation together with an additional non-local interaction.
This non-local interaction captures the dominant effects after microscopic degrees of freedom have been coarse-grained.
Similar to a Ginzburg-Landau approach to phase transitions, we construct a generic model for phase separation with non-local interactions and show that coarsening is generally suppressed.
In addition to this common feature, we also reveal fundamental differences between long-range and short-range interactions, which allow us to categorize the various complex interactions mentioned above.

\mysection{Results}
\label{sec:results}
To develop our theory, we consider an isothermal system of volume~$V$ described by a coarse-grained density field $\phi(\bs x)$ of a single component.
To analyze the effects of non-local interactions, we study the free energy
\begin{multline}
    F[\phi] = \int \diff \bs x  \left \{ f[\phi(\bs x)] + \frac{\kappa h^2}{2} \bigl| \nabla\phi(\bs x) \bigr|^2 \right \}
\\
    + \frac{1}{2}\iint \diff\bs x \, \diff\bs x' g[\phi(\bs x)] K(|\bs x-\bs x'|) g[\phi(\bs x')]
    \;,
    \label{eq:freeEnergy}
\end{multline}
where $f$ accounts for purely local interactions driving phase separation.
The second term on the first line associates an energy penalty~$\kappa$ with compositional variations, resulting in  interfaces of typical width~$h$ between regions of different density.
The second line represents non-local interactions arising from coarse-graining microscopic degrees of freedom  not resolved at our field theoretic level.
We  characterise these interactions by an isotropic kernel $K(|\boldsymbol x-\boldsymbol x'|)$ that sets the interaction strength between the density field at position $\boldsymbol x$ and $\boldsymbol x'$ after a general weight function $g[\phi]$ has been applied.
For simplicity, we here focus on linear weight functions, $g[\phi]=\phi-\bar\phi$ with mean density $\bar\phi=V^{-1}\int \phi \diff V$, which represent pairwise non-local interactions.
Since we consider mass-conserving systems, we employ Model B dynamics~\cite{Hohenberg1977},
\begin{equation}
    \partial_t \phi(\bs x) = \nabla\cdot \left [ \Lambda(\phi)\nabla \frac{\delta F}{\delta\phi(\boldsymbol x)}\right]
    \;,
    \label{eq:modelB}
\end{equation}
where spatial fluxes are driven by gradients in chemical potential $\delta F/\delta\phi(\boldsymbol x)$, and are proportional to the density-dependent mobility $\Lambda(\phi)$.

\mysubsection{Non-local interactions modify stability conditions}
We start by analyzing the stability of homogeneous states for general non-local interaction kernels~$K$.
Inserting \Eqref{eq:freeEnergy} into \Eqref{eq:modelB} and using the perturbative expansion $\phi(\bs x, t) = \bar\phi + \int \diff\bs{q} \hat \phi(\bs q, t)e^{\omega t-i \bs{q}\bs{x}}$, we find the dispersion relation
\begin{align}
   \omega(q) &= -q^2 \Lambda(\bar\phi)\left \{  f''(\bar\phi) +h^2\kappa q^2  +  \hat K(q)\right \}
    \;,
    \label{eq:dispRel}
\end{align}
with wave number $q=|\boldsymbol q|$ (due to isotropy) and Fourier-transformed interaction kernel $\hat K=\int e^{-i\bs{q}\bs x}K(\bs x)\diff\bs x$.
Note that here, and below, primes denote derivatives with respect to the argument of the function.

To understand the influence of non-local interactions, we first ask whether the homogeneous state is stable.
This can be determined from the fastest growing Fourier mode $q_*$, for which $\omega'(q_*)=0$. 
If $\omega(q_*)>0$, the system is unstable, and thus in the \emph{spinodal region}.
Analogous to systems without non-local interactions~\cite{kardar_statistical_2007}, increasing $f''$ (e.g., by increasing the temperature) weakens the instability until the stability boundary is reached at the critical value $\fs''$, i.e., when $\omega(q_*)=0$.
This boundary is thus characterized by the value of $\fs''$ and the corresponding most unstable mode $\qs$, which together satisfy $\omega(\qs)=\omega'(\qs)=0$, such that
\begin{subequations}\label{eq:spinodal}
\begin{align}
    \fs''(\bar\phi) + h^2\kappa \qs^2 +  \hat K(\qs) &= 0 
    \qquad \text{and}
    \label{eq:spin_a}
\\
    2h^2\kappa \qs  + \hat K'(\qs)  &=  0 \label{eq:spin_b}
    \;
\end{align}
\end{subequations}
defines the spinodal boundary.
Consequently, phase separation will occur via spinodal decomposition whenever $f''<\fs''$, e.g., for low enough temperatures.
The associated spinodal mode $\qs$ is equivalent to the marginal mode of the free energy~$F$ (obeying $\partial_q (\delta^2 F/\delta\phi^2)|_{q=\qs} = 0$) since the dynamics given by \Eqref{eq:modelB} minimize the free energy. 
Consequently, the marginal mode $\qs$ characterizes equilibrium and encodes information about the long-term behavior of the system. %
In contrast, the fastest growing mode $q_*$, which we derive in \Appref{app:qstar}, does depend on dynamics and is only relevant for the initial behavior.

Without non-local interactions ($\hat K(q) =0$), \Eqref{eq:spin_b} implies a vanishing marginal mode ($\qs=0$), so %
 modes with arbitrarily large length scales dominate at equilibrium.
Additionally, \Eqref{eq:spin_a} implies that 
marginal stability is achieved when $f''(\bar\phi)=0$, consistent with the classical spinodal analysis of systems without non-local interactions~\cite{kardar_statistical_2007}.
Within the spinodal region ($f''(\bar\phi)<0$), we have $\omega'(q) > 0$, so the system can always be made more stable by increasing length scales (decreasing $q$), leading to coarsening via Oswald ripening. %

Non-local interactions affect the spinodal conditions fundamentally.
First, non-local interactions modify the stability condition \eqref{eq:spin_b}, so the marginally stable mode $\qs$ is typically non-zero, suggesting modes with finite length scale will prevail. %
Second, \Eqref{eq:spin_a} shows that non-local interactions shift $\fs''$ with respect to the case without non-local interactions. 
In particular, non-local interactions destabilize the system and enlarge the spinodal region for $\hat K(\qs) < 0$, so that weaker attractive forces are now sufficient to drive phase separation.
In contrast, \mbox{$\hat K(\qs) > 0$} stabilizes the system, so stronger attractive interactions are needed for phase separation.

To investigate the effect of non-local interactions further, we note that two general classes of kernels can be identified based on whether the spinodal exhibits macroscopic modes or not. %
Macroscopic modes require that $\qs=0$ is a solution to \Eqsref{eq:spinodal}, implying $\hat K'(\qs)=0$.
Consequently, the Fourier-transformed kernel must be differentiable at $q=0$.
This requirement immediately excludes kernels that decay algebraically for large distance in real space since the Paley–Wiener theorem implies that the Fourier transform of such functions is not analytic in the vicinity of $q=0$~\cite{paleyWiener}.
In contrast, analytic kernels that decay exponentially in real space exhibit an analytic Fourier transform, which implies $\hat K'(0)=0$ since $\hat K(q) = \hat K(-q)$.
In this case, $q=0$ is always a solution to \Eqsref{eq:spinodal} for a particular $\fs''$, although $\omega(\qs)$ is not necessarily a maximum and thus the most unstable mode.
In summary, kernels that decay algebraically (long-range interactions) cannot exhibit macroscopically large modes at the spinodal boundary, suggesting that indefinite coarsening is impossible.
In contrast, exponentially decaying kernels (short-range interactions) can in principle exhibit long-term coarsening.
To understand the influence of non-local interactions in more detail, we next discuss particular examples of these two classes.

\mysubsection{Long-range interactions yield regular patterns}
\label{sec:long}
We first study long-range interactions, whose interaction kernels fall off algebraically in space.
Examples of such interactions include electrostatics~\cite{Muratov2002}, the Ohta-Kawasaki free energy~\cite{Ohta1986,Liu1989}, or single molecule chemical reactions with mass-action kinetics~\cite{Christensen1996,Zwicker2022}.
These interactions can all be described by the generic form $\hat K_{\rm{lr}}(q)=\epsilon/q^n$, 
where $\epsilon>0$ sets the energy scale and $n\geq d$ in $d$ spatial dimensions.
Such long-range interactions require additional constraints for the existence of equilibrium states~\cite{Levin2014}, e.g., overall charge neutrality for Coulomb interactions, since otherwise the free energy density becomes extensive with system size. %
The general condition $\int \diff \bs x g[\bar\phi] = 0$ is satisfied by our choice $g[\phi]=\phi-\bar\phi$, but it forbids strictly positive or negative $g[\phi]$, e.g., for gravitational forces. 
Equilibrium states with long-range interactions generally involve screening~\cite{Levin2014}, i.e., a positive $g[\bar\phi]$ region shelled by a negative $g[\bar\phi]$ region, resulting in regular patterns arranged in a hexagonal grid (in $d=2$).
The typical pattern length scale is governed by the dominant spinodal mode~$\qs = \left ( n\epsilon / 2h^2\kappa \right )^{1/(n+2)}$ following from \Eqref{eq:spin_b}.
Inserting $\qs$ into \Eqref{eq:spin_a} yields the spinodal condition $\fs'' = -h^2\kappa\qs^2(n+2)/n$ 
showing that long-range interactions always stabilize homogeneous states since $\fs''<0$. %
In summary, long-range interactions require stronger attractive forces to induce phase separation, and once patterns form they arrest at a finite length scale due to screening. %

\mysubsection{Short-range interactions feature a transition from coexistence of homogeneous phases to patterns}
\label{sec:short}

\begin{figure*}
\begin{center}
\includegraphics[width=2\columnwidth]{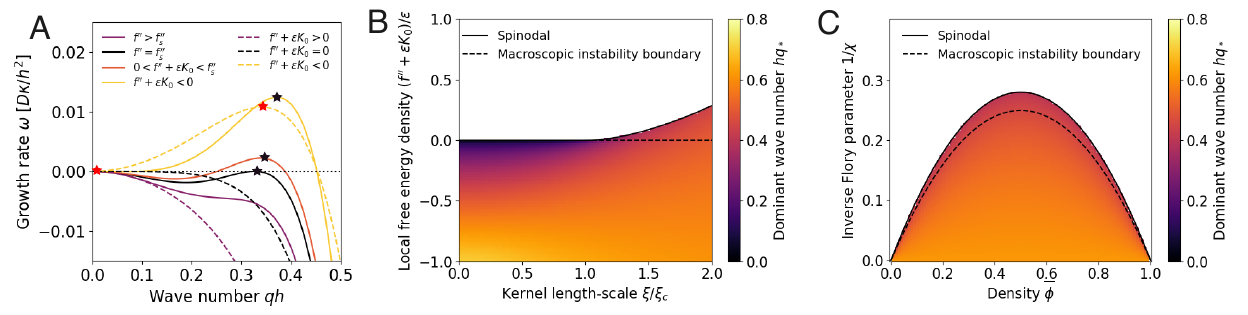} 
\caption{ \textbf{Short-range interactions induce transition from homogeneous phase coexistence to regular patterns.}
(A) Dispersion relation $\omega(q)$ as a function of wave number $q$ for $\xi<\xiC \equiv h\sqrt{\kappa/\epsilon}$ (dashed) and $\xi>\xiC$ (solid) for various $f''$ putting the system outside, at, and inside the spinodal boundary (colors).
The dominant wave number $q_*$ (stars, black for $\xi>\xiC$ and red otherwise) is finite inside the spinodal region; at the spinodal boundary (black lines) $q_*=\qs$ continuously increases from $\qs=0$ to $\qs\neq 0$ as $\xi$ crosses $\xiC$. 
(B) Dominant wave number $q_*$ (Appendix~\ref{app:qstar}) as a function of $\xi/\xiC$ and $f''$ for fixed $\lambda/h=1.4$.
Along the spinodal (solid black line) $\qs$ transitions continuously from $\qs=0$ to $\qs>0$ at $\xi=\xiC$. 
Macroscopic modes with $q\to 0$ are unstable below the dashed line, corresponding to the yellow lines in (A). (C) $q_*$ as a function of overall density $\bar\phi$ and interaction strength $\chi$ for $\xi/\xiC = 1.6$.
(A--C) Model parameters are $\kappa=\chi$, $\lambda/h=1.4$, with $f$ given by \Eqref{eq:FloryF}.
 }
\label{fig:qDyn_f0}
\end{center}
\end{figure*}

We next study short-range interactions, which are described by kernels~$K_\mathrm{sr}(|\boldsymbol{x}|)$ that decay exponentially for large distance.
Examples include non-local elasticity~\cite{Onuki2001-kz, Qiang2024-li, paulin2025}, viscoelastic fluids~\cite{Tanaka2022-la}, bilayer membranes~\cite{Yu2025-eo}, and other forms of differential forces~\cite{Tateno2021-oy}.
Since the Fourier-transform of such interaction kernels is analytic~\cite{paleyWiener}, we describe them by a generic Taylor expansion around $q=0$,
\begin{equation}
	\hat K_\mathrm{sr}(q) \simeq \epsilon\left [K_0 + \xi^2 q^2\left ( -1 + \lambda^2 q^2 \right )\right ]
	\;,
\label{eq:kerAn}
\end{equation}
where we truncate the expansion at fourth order of $q$. %
The length scales $\xi$ and $\lambda$ characterize spatial dependencies of the interaction, 
$\epsilon$ sets the energy scale, and the constant $K_0$ controls whether non-local interactions are overall attractive ($K_0<0$) or repulsive ($K_0>0$).
We consider kernels that decay at intermediate distances ($\xi>0$) and are positive for $q\to\infty$ to stabilize homogeneous regions ($\lambda>0$). %

To see whether short-range interactions exhibit finite pattern sizes, we analyze the spinodal mode $\qs$ by inserting~\Eqref{eq:kerAn} into~\Eqref{eq:spinodal}.
We find that \Eqref{eq:spin_b} always permits the solution $\qs=0$, but $\qs\neq0$ is also possible.
The stability of the system is controlled by the mode with the largest growth rate~$\omega$, so 
we insert both solutions into \Eqref{eq:spin_a} and determine the one with maximal $\fs''$.
This approach yields the spinodal conditions
\begin{equation}\label{eq:spinAn}
\fs'' = \epsilon\left [ - K_0 + \xi^2\lambda^2 \qs^4\right ]
;\;\;
\qs^2 =
\begin{cases}
 \dfrac{\xi^2-\xiC^2}{2\lambda^2\xi^2} & \xi>\xiC \\
 0 & \xi \leq \xiC
 \;,
\end{cases}
\end{equation}
where $\xiC=h\sqrt{\kappa/\epsilon}$ emerges as a critical length, which  %
reveals a competition between the short-range kernel and interface properties.
For $\xi\le\xiC$, we find $\qs=0$, implying that $F$ has a (local) minimum for $q=0$, so
coarsening will always stabilize the system and macroscopic modes ($q\to 0$) dominate, similar to the case without non-local interactions.
In contrast, for $\xi>\xiC$ the free energy $F$ exhibits a minimum at finite $\qs$, suggesting that coarsening is suppressed. 
The transition at $\xi=\xiC$ is continuous since $\qs$ increases continuously from $\qs=0$.
In summary, we find that systems with short-range interactions exhibit a finite spinodal mode $\qs$ for sufficiently large $\xi$.

To learn more about the two qualitatively different regimes identified in \Eqref{eq:spinAn}, we next analyze the dispersion relation $\omega(q)$ given by \Eqref{eq:dispRel}.
We first consider the case  $\xi<\xiC$, where  stability is governed by macroscopic modes with $q\to 0$ and the spinodal is given by the condition $\omega'(0)=0$ (dashed lines in~\Figref{fig:qDyn_f0}A).
Inside the spinodal region, %
the fastest growing mode is finite, but arbitrarily small modes ($q\to 0$) are unstable, so coarsening will take place, similar to systems without non-local interactions. 

In the contrasting case $\xi>\xiC$, stability is governed by the finite mode $\qs$ given by \Eqref{eq:spinAn}.
This finite mode corresponds to a local maximum in the dispersion relation $\omega(q)$, which is already present even when the system is stable (violet line in \Figref{fig:qDyn_f0}A).
The length scale corresponding to this local maximum might be detectable in the fluctuation spectrum of this stable system.
For $f''<\fs''$ the system becomes unstable ($\omega(q_*)>0$), but if $f''>\epsilon K_0$ macroscopic modes are still stable ($\omega'(0)<0$).
This implies a band of unstable modes (orange line), suggesting that patterns with finite length scale form.
In contrast, macroscopic modes become unstable ($\omega'(0)>0$) for $f''<\epsilon K_0$, implying that coarsening could take place (yellow line).
However, the spinodal mode $\qs$ is still finite, suggesting equilibrium states with patterns of a finite length scale since this mode corresponds to the free energy minimum.
While the transition at $\xi=\xiC$ qualitatively influence the spinodal mode, it does not affect the dominant wave number $q_*$  (\Figref{fig:qDyn_f0}B).
We therefore hypothesize that the effects of the transition at $\xiC$ will become evident only at late times. %

\mysubsection{Short-range interactions show rich phase behavior}
\label{sec:SH}

\begin{figure*}
\begin{center}
\includegraphics[trim={0.0cm 0cm 0.0cm 0cm}, clip,width=2.0\columnwidth]{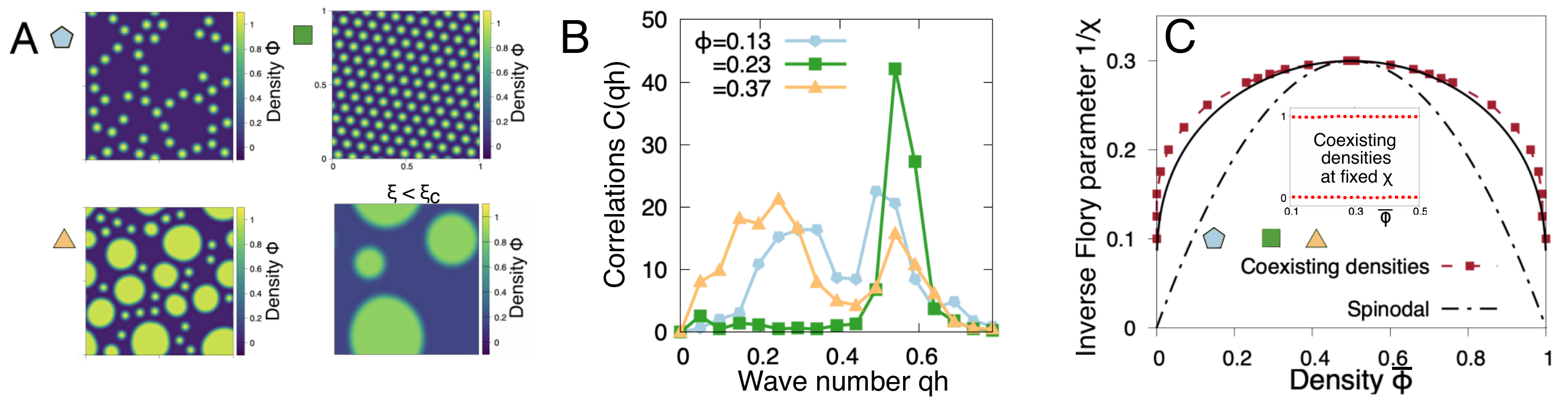} 
\caption{\textbf{Patterned phase can coexist with homogeneous phase for short-range interactions.} %
Density fields $\phi(\boldsymbol{x})$ from 2D numerical integration  (\Appref{app:simulations}) of \Eqsref{eq:freeEnergy}--\eqref{eq:modelB} with short-range interactions given by \Eqref{eq:kerAn} 
for $\xi>\xiC$ (first three panels) and $\xi<\xiC$ (lower right panel).
(B) Density correlations in Fourier space %
for different wave numbers $q$ for each of the three snapshots in (A) with $\xi>\xiC$.
(C) Coexisting densities (red symbols) as function of $\bar\phi$ and $\chi$ are approximated by effective binodal (black line, without non-local interactions), indicating a first order phase transition in two-dimensions.
Inset shows that changing $\bar\phi$ at fixed $\chi=0.1$ does not affect coexisting densities.
Symbols indicate parameter values in panel (A).
(A--C) Parameters are as in~\Figref{fig:qDyn_f0}, except for the lower right snapshot in (A) which has $\xi/\xiC=0.9$.
}
\label{fig:snapsShot}
\end{center}
\end{figure*}

To assess our hypothesis and  test the influence of short-range interactions on the late-time behavior in more detail, we next investigate equilibrium configurations for a particular choice of free energy.
We choose a Flory-Huggins form of the local free energy~\cite{Flory1942,Huggins1941}, %
\begin{equation}
    f[\phi] = \frac{\kB T}{\nu}\left[
        \phi\ln\phi + (1-\phi)\ln(1-\phi) - \frac{\chi}{2}\phi^2
    \right]\;,
    \label{eq:FloryF}
\end{equation}
as it is commonly used in the context of phase separation. The first two terms on the right represent translational entropy while the term proportional to the non-dimensional parameter $\chi$ accounts for pairwise interactions with $\kB T$ the thermal energy and $\nu$ the relevant molecular volume. 
In addition, we consider the case where the short-range kernel given by \Eqref{eq:kerAn} vanishes for large $q$ for simplicity ($K_0=0$).
In this case, the overall effect of short-range interactions is to destabilize the system and shift the spinodal to higher values of $1/\chi$ (\figref{fig:qDyn_f0}C), implying phase separation takes place at weaker attractive interactions $\chi$ when short-range non-local interactions are present.

To go beyond the linear regime, we numerically evolve the dynamics given by \Eqsref{eq:modelB}, \eqref{eq:kerAn}, and \eqref{eq:FloryF} (\Appref{app:simulations}).
\Figref{fig:snapsShot}A shows that full coarsening only takes place for $\xi<\xiC$ (lower right panel), whereas all examples with $\xi>\xiC$ exhibit somewhat regular patterns, consistent with the finite spinodal mode $\qs$ in this case.
However, these numerical results indicate that patterns often contain more than one dominant length scale:
At low overall density  $\bar\phi$ (upper left panel in \figref{fig:snapsShot}A) we observe the coexistence of a dilute homogeneous phase with patterns. 
For slightly larger $\bar\phi$, we observe a pure patterned phase with one dominant length scale (upper right panel).
At even larger $\bar\phi$, the system exhibits coexistence of dense homogeneous phases with patterns.
These visually apparent transitions are corroborated by the correlations of the density fields, which exhibit two peaks for the complex patterns, whereas the pure patterned phase exhibits a single peak (\Figref{fig:snapsShot}B).
Taken together, equilibrium states with short-ranged interactions exhibit patterned regions for $\xi>\xiC$, which can sometimes coexist with homogeneous phases.

To characterize the patterns further, we next analyze the average densities of low-density and high-density regions (see Appendix~\ref{app:simulations} for details).
The inset of \Figref{fig:snapsShot}C shows that these densities are independent of the total density $\bar\phi$, similar to coexisting densities in phase separation without non-local interactions. 
This is surprising since individual droplets in the patterned phase are not thermodynamic phases, so arguments from the thermodynamics of phase coexistence need not apply.
Rather, all droplets together form the thermodynamic patterned phase and only the average density in this phase is thermodynamically constrained.
Inspired by this observation, we next ask how measured densities vary with the interaction parameter $\chi$ at fixed overall density $\bar\phi$.
\Figref{fig:snapsShot}C shows that measured densities get closer to $0.5$ for larger $\chi^{-1}$, similar to coexisting densities in systems without non-local interactions.
In fact, the family of densities (red dots) is well described by the binodal (black line) of an effective system without non-local interactions, but with a shifted interaction parameter $\tilde\chi=\chi+\fs''$. This observation is consistent with our previous conclusion that non-local short-range interactions can promote  ($\fs''>0$) or suppress ($\fs''<0$) 
phase separation.
This analysis suggests that the system exhibits a first-order phase transition, indicating that suddenly a patterned phase with finite amplitude emerges when $\chi$ is increased (except at the critical point). %

\mysubsection{Short-range interactions can be mapped to conserved Swift--Hohenberg model}
\label{sec:SH}

The phases we observe for $\xi>\xiC$ are similar to those observed in the conserved Swift--Hohenberg model, e.g., used to describe crystal growth~\cite{BuKn2006pre,  MaBK2010pd, HAGK2021ijam}.
In fact, the free energy given by \Eqref{eq:freeEnergy} with the generic short-range kernel given by \Eqref{eq:kerAn} can be mapped directly to a Swift--Hohenberg-like free energy (Appendix~\ref{app:swift_hohenberg}),
\begin{multline}
   F_{\rm SH}  = \int \diff \bs x  \left \{ f[\phi(\bs x)] - \fs''\frac{\phi^2(\bs x)}{2} \right \}\\
+ \frac{\epsilon\xi^2\lambda^2}{2}\int \diff \bs x\left \{ \phi(\bs x)\left ( \nabla^2 + \qs^2 \right )^2 \phi(\bs x)\right \} 
    \;,
   \label{eq:swiftHoh}
\end{multline}
where we have assumed that $\xi>\xiC$, implying $\qs>0$. 
Consequently, modes with $q=\qs$ minimize the non-local part of the free energy (second line), suggesting that patterns with such a length scale could be stable.
However, the detailed equilibria depend on all terms in \Eqref{eq:swiftHoh}, and thus the precise form of the local free energy $f$.
 
The minima of $F_{\rm{SH}}$ have been studied for the case where $f[\phi]$ is a polynomial in $\phi$, e.g., in the context of phase-field models of crystal growth~\cite{BuKn2006pre,  MaBK2010pd, HAGK2021ijam}. %
Refs.~\cite{Emmerich2012,Thiele2013,BuKn2006pre,  MaBK2010pd, HAGK2021ijam} contain detailed calculations of the corresponding phase diagram, which shows the same qualitative behavior described above, including coexistence of homogeneous phases and patterns.
Moreover, Ref.~\cite{Thiele2013} also shows that fixing the density $\bar\phi$ but changing a control parameter analogous to $1/\chi$ leads to a continuous phase transition in one spatial dimension, but a discontinuous transition in two or higher dimensions, the latter being consistent with our observations in two dimensions.  
The similarity in the phase diagrams suggests that results from phase-field descriptions of crystal growth can be transferred to models describing phase separation with short-range interactions, where $f$ given by \Eqref{eq:FloryF} is more natural.
In particular, we expect that the amplitude equations derived in Ref.~\cite{Thiele2013} also hold for $f$ given by~\Eqref{eq:FloryF} when $\chi$ and $\bar\phi$ are sufficiently close to the critical point.
Similarly, finite systems should exhibit a series of bifurcations, one for each droplet that is formed/removed from the system as density is changed~\cite{Thiele2013}.
Finally, disturbances in these bifurcations at sufficiently low $1/\chi$ are expected to be directly related to the emergence of macroscopic droplets that we describe here~\cite{MaBK2010pd, HAGK2021ijam}.
Taken together, the mapping to the conserved Swift--Hohenberg model explains similar behavior observed across different systems that couple phase separation with effective short-range interactions~\cite{Onuki2001-kz, Qiang2024-li, Mannattil2024-lf, paulin2025, Tanaka2022-la, Yu2025-eo, Tateno2021-oy}.

\textit{Discussion---}We have explored how non-local interactions affect phase separation.
Similar to classical systems, the homogeneous state is stable if the driving force toward phase separation is weak (large $f''$).
However, if phase separation takes place, non-local interactions affect the system qualitatively.
In particular, long-range (algebraic) interactions always form regular patterns due to screening.
In contrast, short-range interactions only slightly modify the classical behavior if they are  weak ($\epsilon\xi^2 < \kappa h^2$).
However, strong short-range interactions exhibit patterns, which can coexist with homogeneous macroscopic phases, depending on the overall composition.
In this case, the free energy can be mapped to a conserved Swift--Hohenberg model, revealing how different short-range interactions exhibit universal behavior.

Our generic analysis provides a guideline to interpret results from different phase-separating systems with complex interactions.
For instance, the mapping to the Swift--Hohenberg model resolves a discrepancy observed in models of phase separation with non-local elasticity with a Gaussian kernel~\cite{Qiang2024-li,Mannattil2024-lf}.
While Ref.~\cite{Mannattil2024-lf} found a first-order transition in three dimensions, Ref.~\cite{Qiang2024-li} discovered a continuous transition in one dimension, both consistent with the Swift--Hohenberg model.
In contrast, experiments suggest a continuous phase transition in three-dimensional elastic gels~\cite{Fernandez-Rico2022-wu,Fernandez-Rico2024-ap}, suggesting that neither theory captures all relevant details.
More generally, we expect that our results provide a suitable classification for many complex phase separating systems.

\textit{Acknowledgements---}%
We thank Uwe Thiele, Riccardo Rossetto, and Guido Kusters for helpful discussions, and Chengjie Luo for critical comments on the manuscript. 
We gratefully acknowledge funding from
the Max Planck Society and the European Union (ERC, EmulSim, 101044662).

\bibliographystyle{apsrev4-1}

%

\newpage 

\onecolumngrid

\appendix
\setcounter{figure}{0}
\renewcommand{\figurename}{Fig.}
\renewcommand{\thefigure}{S\arabic{figure}}

\section{Fastest growing mode}
\label{app:qstar}
A necessary condition for the fastest growing mode $q_*$ is $\omega'(q_*)=0$, where $\omega(q)$ is given by \Eqref{eq:dispRel}.
Hence,
\begin{equation}
	\label{eq:qstar}
    q_*\left (  f''(\bar\phi) + 2h^2\kappa q^2_*  +    \hat K(q_*) + \frac{q_*}{2} \hat K'(q_*) \right ) =  0
    \;.
\end{equation}
For the short-range interactions given by \Eqref{eq:kerAn}, this becomes
\begin{equation}
	q^2_* = 
    \dfrac{\xi^2-\xiC^2}{3\lambda^2 \xi^2}\left [ 1 \pm 
    \sqrt{1 - \dfrac{3(f''+\epsilon K_0) \lambda^2 \xi^2 }{\epsilon (\xi^2-\xiC^2)^2}} \right ]
	\;,
\label{eq:qdynAn}
\end{equation}
which reduces to \Eqref{eq:spinAn} for $\qs$ when $f'' = \fs''$.
Here, the $\pm$ represent the solutions for $\xi>\xiC$ (+) and for $\xi<\xiC$ (-).

\section{Simulation details}
\label{app:simulations}
We integrate \Eqref{eq:modelB} numerically in Fourier space using a pseudo-spectral method~\cite{Zhu1999-hm}.
To avoid numerical problems with the logarithm in the free energy given by \Eqref{eq:FloryF}, we regularize $\ln(\phi)$ as
\begin{equation}
    \widetilde \ln(\phi) = 
    \begin{cases}
        \ln (\phi) & \textrm{if } \phi>\delta\\
        \ln(\delta) - \frac{\phi-\delta}{\delta} +  \frac{(\phi-\delta)^2}{2\delta^2} & \textrm{if } \phi\leq\delta
        \;, \\
    \end{cases}
\end{equation}
To recover Fickian diffusion for $\phi\to 0$, and to preserve symmetry with respect to $\phi=\frac12$, we use the mobility
\begin{equation}
    \Lambda (\phi) = D\left \{ \frac{\partial^2}{\partial \phi^2}\left [ \phi\widetilde \ln(\phi) + (1-\phi)\widetilde \ln(1-\phi) \right ] \right \}^{-1}
    \;,
\end{equation}
where $D$ is a constant.
This recovers the usual expression $\Lambda = D\phi(1-\phi)$ for $\delta < \phi < 1-\delta$ and guarantees that the diffusivity $\Lambda(\phi)f''$ remains constant and equal to $D$ for ideal systems ($\chi=0)$ for all $\phi$.
Similar relations are used in macroscopic fluctuation theory~\cite{Derrida2011-le}, but this aspect is often overlooked in literature on Cahn-Hilliard models with a regularized logarithm, which could result in unphysical discontinuities in the diffusivity. %
We use $\delta=10^{-2}$ in all our simulations, and we verified that lowering $\delta$ to $10^{-4}$ did not change any results significantly.

We extract coexisting densities inside and outside droplets for large times by identifying the two dominant peaks in the density histograms. To do so, we bin the density space $\phi\in(0,1)$ into intervals of width $0.01$ and compute corresponding histogram. We then identify the maxima in the histogram in the low ($\phi<0.5$) and high ($\phi>0.5$) density region, respectively, which is the data reported in \figref{fig:snapsShot}C. In the patterned phase, the area occupied by interfaces between microscopic droplets is comparable to the area occupied by their interiors and, together with the observation that the density profile does not change monotonically as an interface is crossed, our method to determine the coexisting densities leads to small deviations of the true peaks towards slightly higher values (e.g., when compared to the density at the center of the droplet) or slightly lower values (i.e., at the middle point between two microscopic droplets). This deviation is expected, since microscopic droplets do not form a thermodynamic phase, meaning ``interface'' contributions do not vanish for large systems.

\section{Mapping to Swift--Hohenberg free energy}
\label{app:swift_hohenberg}
The expansion of the Fourier-transformed short-range kernel given by \Eqref{eq:kerAn} corresponds to a gradient expansion in real space, $K_\mathrm{sr}(|\bs r|) \simeq \epsilon\left [ K_0\delta(|\bs r|) + \xi^2\nabla^2 + \xi^2\lambda^2\nabla^4\right ]$. 
We thus consider the full non-local free energy given by \Eqref{eq:freeEnergy} for a generic isotropic kernel expanded as $K(|\bs x-\bs x'|) = \epsilon K_0+\sum a_n \nabla^{2n}\delta(\bs x - \bs x')$,
\begin{equation}
    F = \int \diff \bs x  \left \{ f[\phi(\bs x)] + \frac{h^2\kappa}{2} \bigl| \nabla\phi(\bs x) \bigr|^2 \right \} 
    +  \frac{1}{2}\int \diff\bs x  g[\phi(\bs x)] \left [ \epsilon K_0+\sum_{n=1} a_n \nabla^{2n} \right ] g[\phi(\bs x)]
    \;.
\end{equation}
Integrating the $\kappa$ term by parts gives %
\begin{equation}
    F = \int \diff \bs x  \left \{ f[\phi(\bs x)] - \frac{h^2\kappa}{2}\phi\nabla^2\phi \right \} 
    +  \frac{1}{2}\int \diff\bs x  g[\phi(\bs x)] \left [ \epsilon K_0+\sum_{n=1} a_n \nabla^{2n} \right ] g[\phi(\bs x)]
    \;,
\end{equation}
where we assume that boundary terms vanish, e.g., because of periodic boundary conditions or because $g[\phi]$ vanishes at the system boundary.
Truncating at $n=2$ and letting $g[\phi(\bs x)]=\phi(\bs x) - \bar\phi$, 
\begin{equation}
    F = \int \diff \bs x  \left \{ f[\phi(\bs x)] + \frac{\epsilon  K_0\phi^2(\bs x)}{2} \right \}
    +  \int \diff \bs x\left \{ \frac{\phi(\bs x)}{2}\left [ \left ( a_1 - h^2\kappa \right )\nabla^2 + a_2\nabla^{4}\right ] \phi(\bs x)\right \} 
    \;.
    \label{eqApp:freeEn}
\end{equation}
Completing the square in the square bracket and identifying $\qs^2=(a_1-h^2\kappa)/(2a_2)>0$, %
\begin{equation}
\begin{split}
    F_\mathrm{SH}  = \int \diff \bs x  \left \{ f[\phi(\bs x)] + \left (\epsilon K_0-\frac{\qs^4 a_2}{2}\right )\frac{\phi^2(\bs x)}{2} \right \}
    +  \frac{a_2}{2}\int \diff \bs x \phi(\bs x)\left ( \nabla^2 + \qs^2 \right )^2  \phi(\bs x)
    \;,
\end{split}
\end{equation}
gives Eq.~\eqref{eq:swiftHoh} in the main text with $a_1=\epsilon\xi^2$ and $a_2=\epsilon\xi^2\lambda^2$.

\end{document}